# THE LINAC LASER NOTCHER FOR THE FERMILAB BOOSTER*

David E. Johnson#, Kevin Laurence Duel, Matthew Gardner, Todd R. Johnson, David Slimmer (Fermilab, Batavia, Illinois), Sreenivas Patil (PriTel, Inc., Naperville, Illinois), Jason Tafoya (Optical Engines, Inc., Colorado Springs, Colorado)

*Abstract*

In synchrotron machines, the beam extraction is accomplished by a combination of septa and kicker magnets which deflect the beam from an accelerator into another. Ideally the kicker field must rise/fall in between the beam bunches. However, in reality, an intentional beam-free time region (aka "notch") is created on the beam pulse to assure that the beam can be extracted with minimal losses. In the case of the Fermilab Booster, the notch is created in the ring near injection energy by the use of fast kickers which deposit the beam in a shielded collimation region within the accelerator tunnel. With increasing beam power it is desirable to create this notch at the lowest possible energy to minimize activation. The Fermilab Proton Improvement Plan (PIP) initiated an R&D project to build a laser system to create the notch within a linac beam pulse at 750 keV. This talk will describe the concept for the laser notcher and discuss our current status, commissioning results, and future plans.

## MOTIVATION

The current Fermilab Booster utilizes multi-turn injection and adiabatic capture to populate all RF buckets in the ring. To minimize losses from the rise time of the 8 GeV extraction kicker, a portion of the beam (about 60 ns. out of 2.2 μs.) in the ring is removed by fast kickers at low energy into an absorber. This empty section of the circumference is called a “Notch”. On Booster cycles that are ultimately injected into the MI for the Neutrino and Muon programs this process takes place at approx.400 MeV about a hundred microseconds after capture. To keep the activation of tunnel components to reasonable limit for maintenance, the Booster has an administrative limit on loss power of 525W. The process of creating the notch in the Booster contributes approximately 1/3 (175W) to this limit. Moving this process out of the Booster tunnel to the 750 keV Medium Energy Beam Transport (MEBT) of the linac is expected to reduce the loss to ~17 W, assuming a 90% efficiency in the Linac Notch creation and 10% clean-up in the Booster ring.

## LINAC & BOOSTER BEAM

At the completion of PIP, the Fermilab Booster will be operating at 15 Hz. The length of the linac pulse injected into Booster is N*τ where N is the number of injected turns and τ is the Booster revolution period at injection. Creating a notch in the linac pulse requires removing N-1 sections of the linac beam at the Booster revolution period. The spacing between these removed sections should guarantee that when the H- is injected into the Booster, the empty sections fall on top of one another in the ring producing a single notch in the Booster. This process is shown in Fig. 1. The top pane shows the 15Hz linac pulses to be injected into Booster. The bottom pane shows a single linac pulse with 60 ns notches created within the pulse separated by ~2.2 μs, the Booster revolution period. Not shown is the 200 MHz bunch structure in the linac pulse.

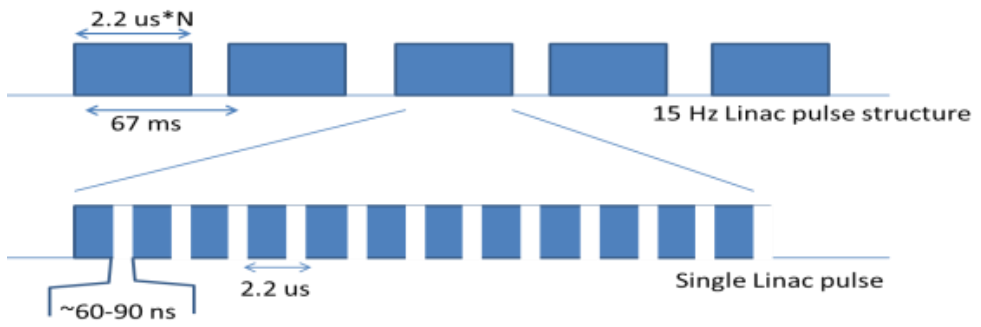

Figure 1: Linac pulse showing the notch structure for a single linac pulse.

## LASER NOTCHER CONCEPT

The technique employed to produce the notch is to remove the outer electron of the H- ion using photoionization for the appropriate beam sections. There have been discussions on using lasers to create a notch in the linac beam for some time [1-3]. This technique was demonstrated in 2000. [1] The photoionization cross section has a broad peak centered at 1.51 eV (λ=821 nm) photon energy in the center-of-mass frame of the electron with a cross section of $4.2\times10^{-17}$ $cm^2$. [4] The choice of the lab frame photon energy is dependent of the H- energy and the interaction angle through the standard Lorentz transformation. The laser technology for both solid state (Nd:YAG) and ytterbium (Yb) doped fiber with a laser wavelength of 1064 nm has matured significantly over the last decade such that it is the natural choice for the laser system. The cross section for these photons with CM energy 1.165 eV is $3.66\times10^{-17}$ $cm^2$, only 13% off the peak.

When the probability of interaction between the photons and electrons is high and the mechanism does not depend on the electron intensity [4], the fraction of electrons that are detached from the moving H- ions is given by [5]

$$F_{neut} = {}^{N}\!/_{N_0} = (1 - e^{-f_{CM}\sigma(E)\tau}), \quad (1)$$

where $f_{CM}$ is the flux of photons at the interaction point in the rest frame of the H- [photons/cm$^2$/sec], σ(E) is the photoionization cross section for photon energy E, and τ is the interaction time of the photons and electrons. The center of mass flux can be expressed in lab frame parameters as

$$f_{CM} = \gamma\left(\frac{E_{laser}\lambda_{LAB}}{hc\tau_{laser}}\right)\left(\frac{1}{A_{laser}}\right)(1-\beta\cos\theta), \quad (2)$$

where $\boldsymbol{E}_{\mathbf{laser}}$ is the laser pulse energy, $\boldsymbol{\lambda}_{\mathbf{LAB}}$ is the lab frame wavelength of the laser, $\tau_{laser}$ is the laser pulse length, $\boldsymbol{A}_{\mathbf{laser}}$ is the laser cross sectional area, γ and β are the usual relativistic parameters, and $\theta$ is the interaction angel between the photons and H-. Figure 2 shows the “single pass” neutralization fraction as a function of laser pulse energy to neutralize a single 60 ns section of the 750 keV H- linac pulse, assuming a 90° interaction angle.

*Operated by Fermi Research Alliance, LLC under contract No. DE-AC02-07CH11359 with the United States Department of Energy.
#DEJ@fnal.gov

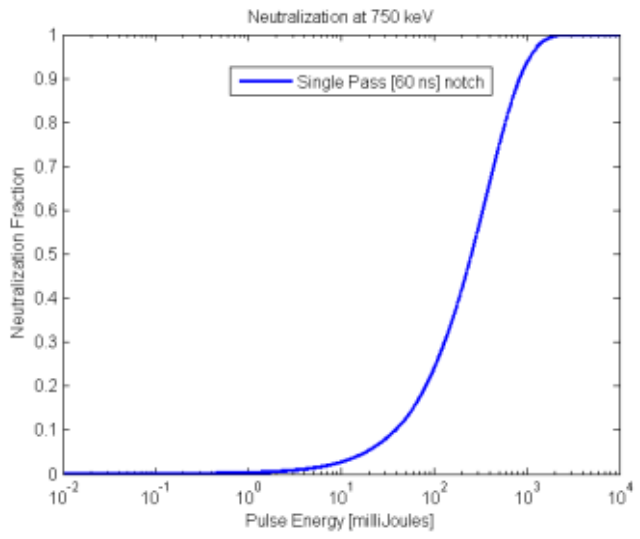

Figure 2: Single pass neutralization fraction for a 60 ns laser pulse.

This shows that to neutralize 95% of the ions would require a laser with ~ 1 Joule pulse energy and 60 ns pulse length which corresponds to a 16 MW/ pulse peak power, a significant laser. To create multiple notches in the linac this laser would have to have a repetition rate of 450 kHz. Such a laser system is beyond what is technically feasible for this project.

## *Reduction of Pulse Energy*

To reduce the laser pulse energy, two techniques are utilized. First, instead of a single pass interaction, we introduce an optical zig-zag cavity [6] made up of two parallel mirrors such that as the laser traverses the cavity it will interact with the ions M times, where M is the number of times the laser crosses the mid- point of the cavity (i.e. ion trajectory). The diameter of this cavity and the angle at which the laser is injected are matched to the ion velocity through the cavity. Second, the laser pulse length is matched to the ion bunch length, otherwise the temporal section of the laser pulse between bunches is wasted and only contributes to the average power of the laser system.

## *Optical Cavity*

The optical cavity is installed in the 750 keV MEBT where the loss of the neutralized ions by the laser do not contribute to activation of any accelerator components. The optical cavity is built into the downstream RFQ flange and just upstream of the first quad in the MEBT. Figure 3 shows a model of the installation with the flange removed to illuminate the geometry of the optical cavity. The end of the RFQ is shown at the bottom and the first quad in the MEBT is at the top of the picture. The cavity is seen between these two devices. The laser enters from the left of the picture, traverses the cavity, and exits on the right. Viewports with an Anti-Reflective coating used to bring the laser into the vacuum cavity are seen on the left and right. The H- traverses the cavity from bottom to top, on the cavity axis. Due to the severe longitudinal space constraints the maximum length of the mirrors in the zig-zag cavity is 25.4 mm which allows for 19 to 23 interactions of the laser with the same group of H- ions. The laser exits the interaction cavity into a laser dump enclosure containing an energy meter for measuring total energy of the laser pulse each 15Hz linac cycle.

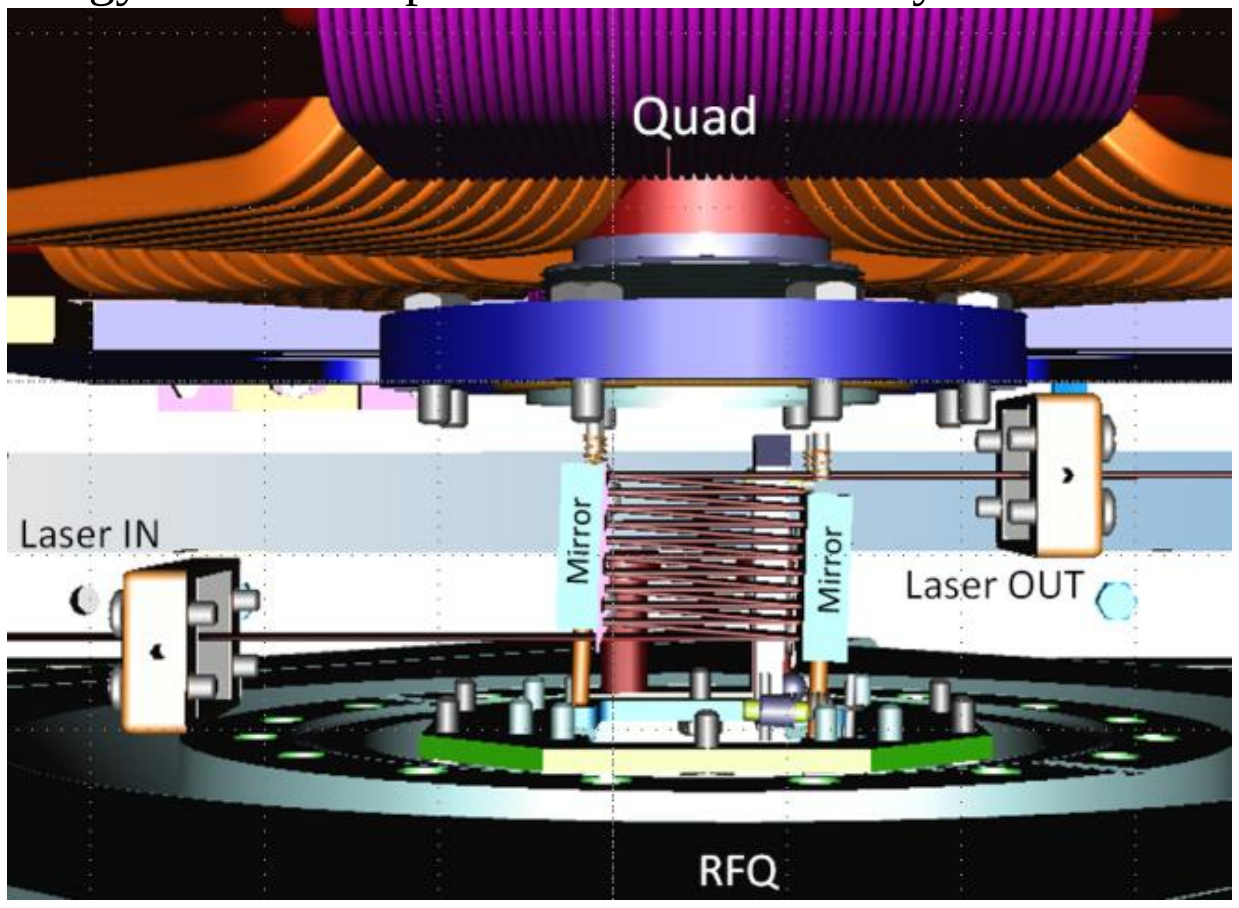

Figure 3: The laser notcher optical cavity between the RFQ and first MEBT quad.

Figure 4 shows the neutralization fraction for a multi-pass cavity (with M=21) and a laser pulse length matched to the ion bunch length (2 ns). This reduces the laser pulse energy from ~1 J/pulse to 2 mJ/pulse *12 pulses/60ns or 24 mJ/60 ns notch. The peak power is reduced from 16 MW to ~ 1 MW.

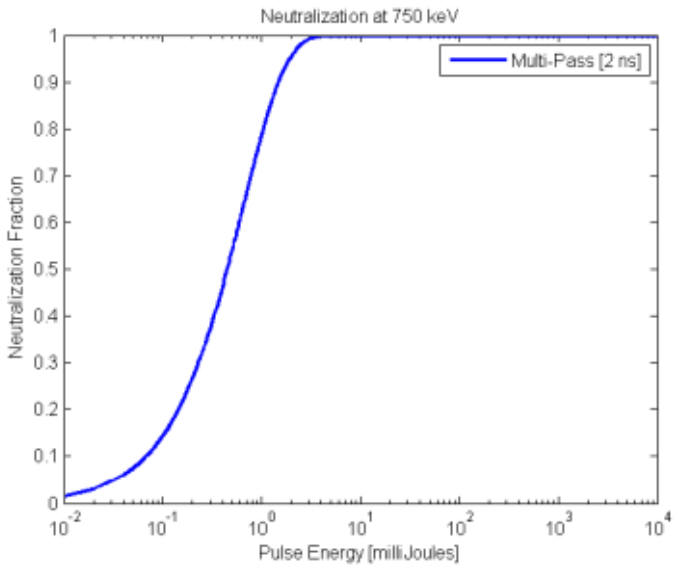

Figure 4: Multi-pass neutralization fraction for a 2 ns laser pulse.

## *Laser Requirements*

The second technique to reduce the laser pulse energy is to match the laser temporal structure to the linac bunch structure out of the RFQ. Figure 5 shows the 200 MHz bunch structure out of the RFQ, the laser pulse structure, and the resultant linac bunch structure to be accelerated in the linac and injected into the Booster.

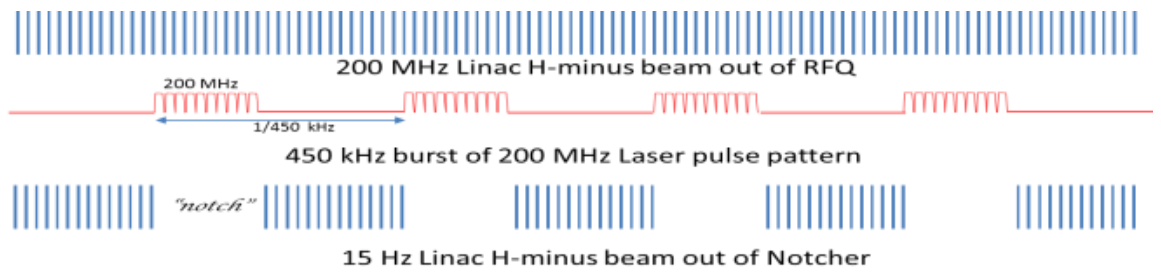

Figure 5: Result of laser pulses on a continuous 200 MHz bunch pattern.

The laser system must be capable of generating a 450 kHz burst of 200 MHz laser pulses each with a pulse length of about 2 ns (equal to the bunch length out of the RFQ) each 15 Hz linac cycle. The minimum number of bursts each 15 Hz is N-1, where N is the number of turns injected into Booster. Since we would like to neutralize nearly 100% of the ion bunch, all ions should see the same photon density thus leading to uniform temporal and spatial profiles. The laser pulses must be synchronized with the 200 MHz bunch structure and the burst must be times so the notches line up once injected into Booster.

## LASER SYSTEM DESIGN

The laser system is a Master Oscillator Power Amplifier (MOPA) configuration which takes a low power seed laser and amplifies it to the required power. Transverse polarization is maintained through the entire laser and optical system. Figure 6 shows a block diagram of the laser and optical system. This depicts the three stage fiber amplifier system followed by a two stage free space amplifier system and beam shaping system to obtain the desired spatial profile (shown at the bottom left) in the interaction cavity. The output of Fiber Amp 2 is brought into free space by a fiber port. All free space amplifiers, optics and transport to the laser dump are contained in interlocked light tight enclosures as this system operates in an area where no protective eyewear is required for people working in the area

Since the laser system must match the temporal structure of the 200 MHz ions exiting the RFQ and the Booster revolution frequency (450 kHz), a CW diode seed laser and a 10 GHz wave-guide modulator are utilized to produce the desired pattern. The modulator is driven by a flexible Arbitrary Waveform Generator (AWG) waveform created in the LabView control software. The software allows control of 200 MHz pulse width, number of pulses per notch, the number of notches, and allow for individual pulse amplitudes within a notch pulse (all equal or increasing or decreasing amplitude, in a linear, quadratic, or cubic function). This pulse pattern is amplified using a three stage fiber amplifier system to increase the pulse energies to ~5μJ. The laser is brought out into free space where it is further amplified by double passing through two diode pumped solid state (DPSS) amplifier modules delivering the desired pulse energy of 2 mJ. The pre-amp and fiber amp 1, built by PriTel Inc., operate CW while fiber amp 2, built by Optical Engines Inc., and both DPSS modules, built by Northrup Grumman, operate at 15 Hz. A system of Faraday isolators and wave plates are used for polarization control to allow double pass operation of the DPSS amplifiers. Several optical telescopes are used to match between the DPSS amplifiers and into an optical system to create a roof-top spatial profile. After exiting the final amplifier, the Gaussian laser profile is focused onto an optical system which splits the single Gaussian beam into 8 beamlets of alternating polarization thus creating a roof-top spatial profile. A final horizontal and vertical cylindrical telescope modify the profile to create a beam with horizontal width of 0.5 to 0.8 mm and vertical height of 10 mm to match the vertical dimension of the ion beam out of the RFQ. The laser then passes through a transport enclosure containing a piezo controlled steering mirror and two optical BPM's which in conjunction with a piezo mirror at the exit of the cylindrical telescopes allow precise control of the laser position and angle into the interaction cavity. The output of the Fiber Amp 2 and both DPSS modules are sampled by a fast photodiode to measure optical pulse structure and a Fermi built integrating sphere to measure total power each linac cycle. At the exit of the cavity an energy meter is installed in the dump enclosure to measure the total energy

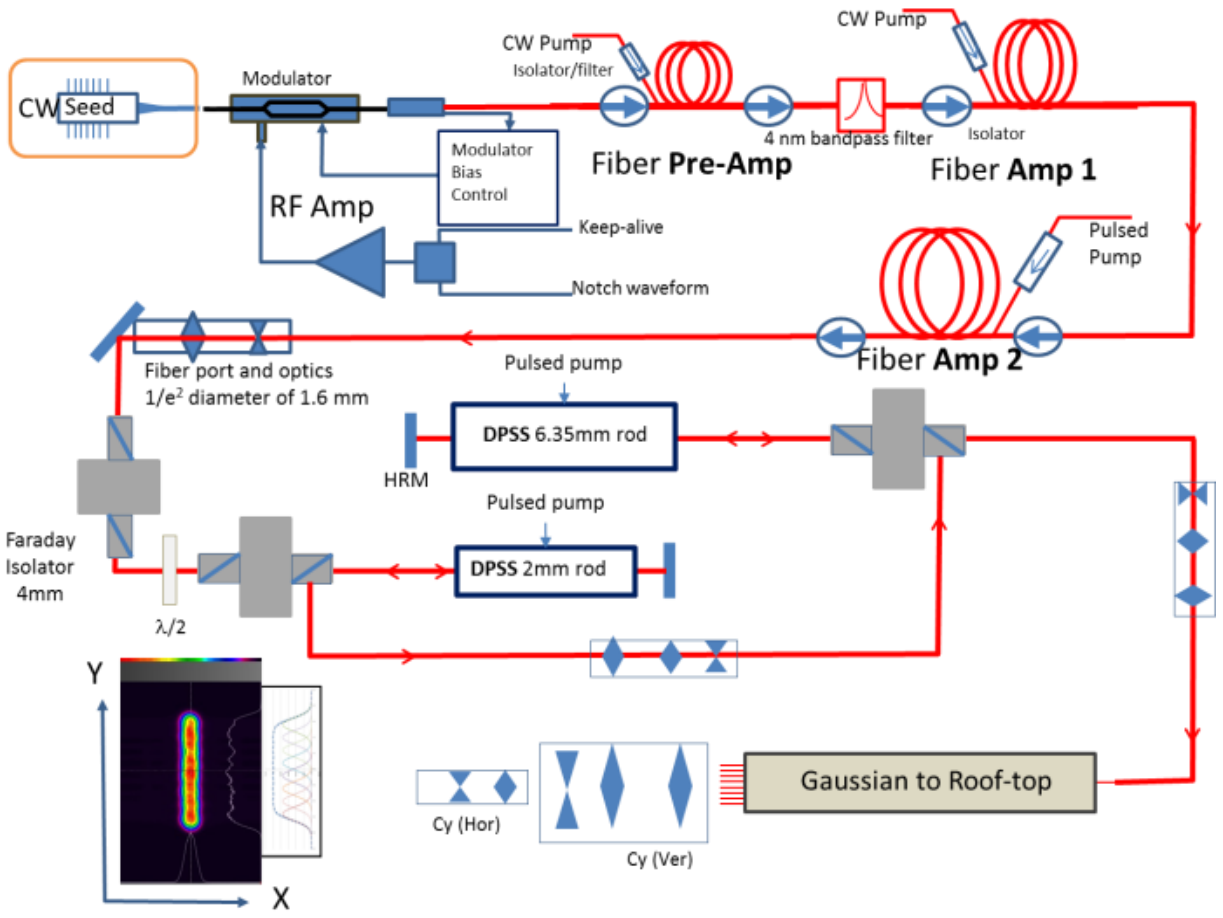

Figure 6: Simplified block diagram of the Linac Notcher laser and optical system components.

## JULY COMMISSIONING SESSION

Although the laser system was not producing the design pulse intensity, it was installed in the Linac prior to the scheduled accelerator shutdown period so that there would be an initial commissioning period to verify the "*as is"* system performance in the electrically noisy and environmentally harsh (as compared to the lab) location next to the RFQ to determine modifications needed during the shutdown period. Once installed, a two week commissioning period with beam took place. Commissioning was done during normal Operation and was parasitic to Operations.

### *Commissioning Goals*

The goals of the commissioning period were 1) verify the laser system could be synchronized to the ion bunch structure exiting the 750 keV RFQ and neutralize multiple sections of 200 MHz bunches in the linac beam pulse separated by the revolution period of the Booster, 2) measure the neutralization efficiency as a function of laser pulse energy, 3) investigate Booster parameters for preservation of the notch during injection, capture, and acceleration, 4) measure the bunch length of the ions exiting the RFQ, 4) gain experience in steering the laser into the cavity, and 5) gain experience with operation in an electrically noisy next to the RFQ. The information learned during this commissioning period will be utilized in the planned improvements during the accelerator shutdown period to bring the pulse energy up to design levels.

## *Observation of multiple notches in Linac pulse*

Several pieces of existing linac instrumentation at various points in the linac were utilized to look at the linac bunch structure. All of the instruments were located either mid-way down the linac or at the end of the linac and consisted of a BPM plate signal in the 400 MeV Booster injection line and a stripline detector and a resistive current wall monitor (RWCM) midway down the linac. Both the BPM and stripline produced doublet signals whereas the RWCM was a singlet signal. The degree of neutralization can be determined by comparing the relative amplitude of the signals for the affected bunches to those adjacent to the notch. Future efforts include fitting of the RWCM signal to determine the degree on neutralization. The results from the three detectors were in reasonable agreement. Due to the bunch-by-bunch variation in intensity throughout the linac pulse, the data were averaged over multiple linac cycles, typically 32 to 64.

The first task after the installation and alignment of the free space optical system enclosures was to steer the beam into the optical cavity to maximize the transmission thru the cavity. Due to the limited study time, an efficiency of 75-80% was deemed acceptable. Once beam operations ended additional time was devoted to steering where ~90% was achieved and additional procedures for steering were developed.

Once the laser system was timed in to match the 200 MHz ion bunch structure out of the RFQ, the linac pulse structure showed for the first time very clean notches at the Booster revolution period for the entire linac pulse. Figure 7 shows the linac pulse for 15 turn injection (~33.27 us).

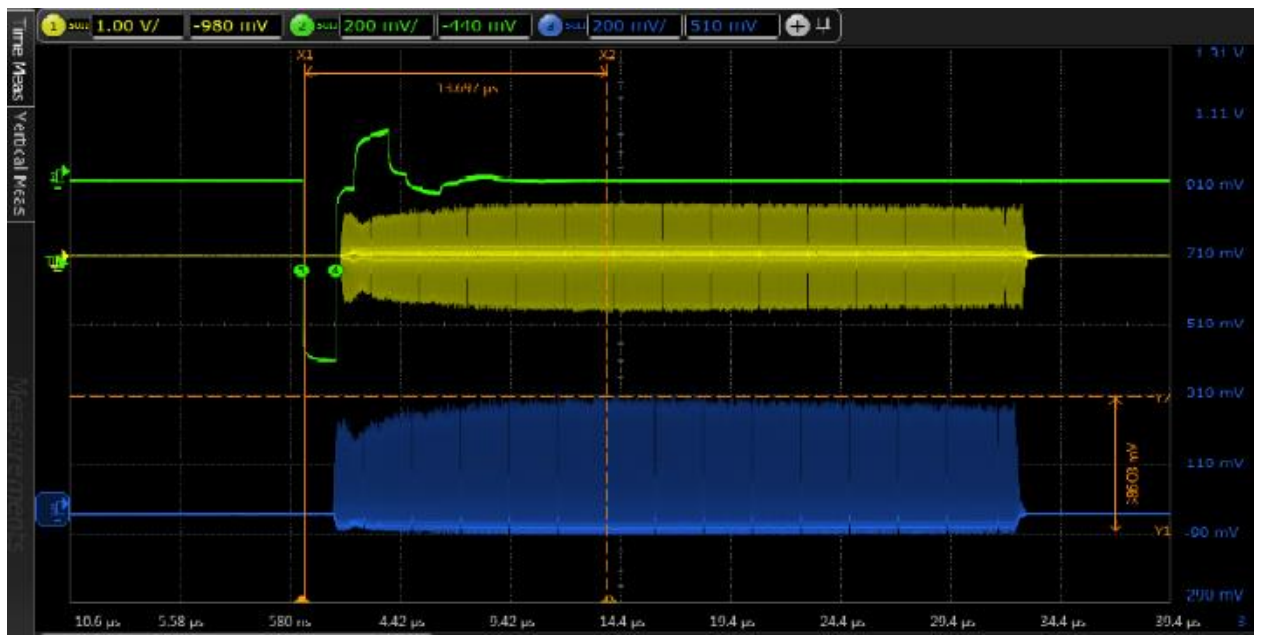

Figure 7: The linac pulse as seen on a stripline detector (yellow) and RWCM detector (blue). The neutralized section of the linac pulse (notches) separated by the Booster revolution period can be seen. The laser trigger (green) can be seen.

Figure 8 shows the bunch intensity of a single notch on two different detectors. The data shown in figures 7 and 8 are taken with a pulse energy of about 0.5 mJ per 200 MHz pulse. To determine the average pulse energy of the 200 MHz laser pulses, the energy meter in the laser dump measured total energy of the laser pulse each linac cycle. The data from the photodiode monitoring the pulse structure out of the final amplifier was used to determine the individual 200 MHz pulse energies.

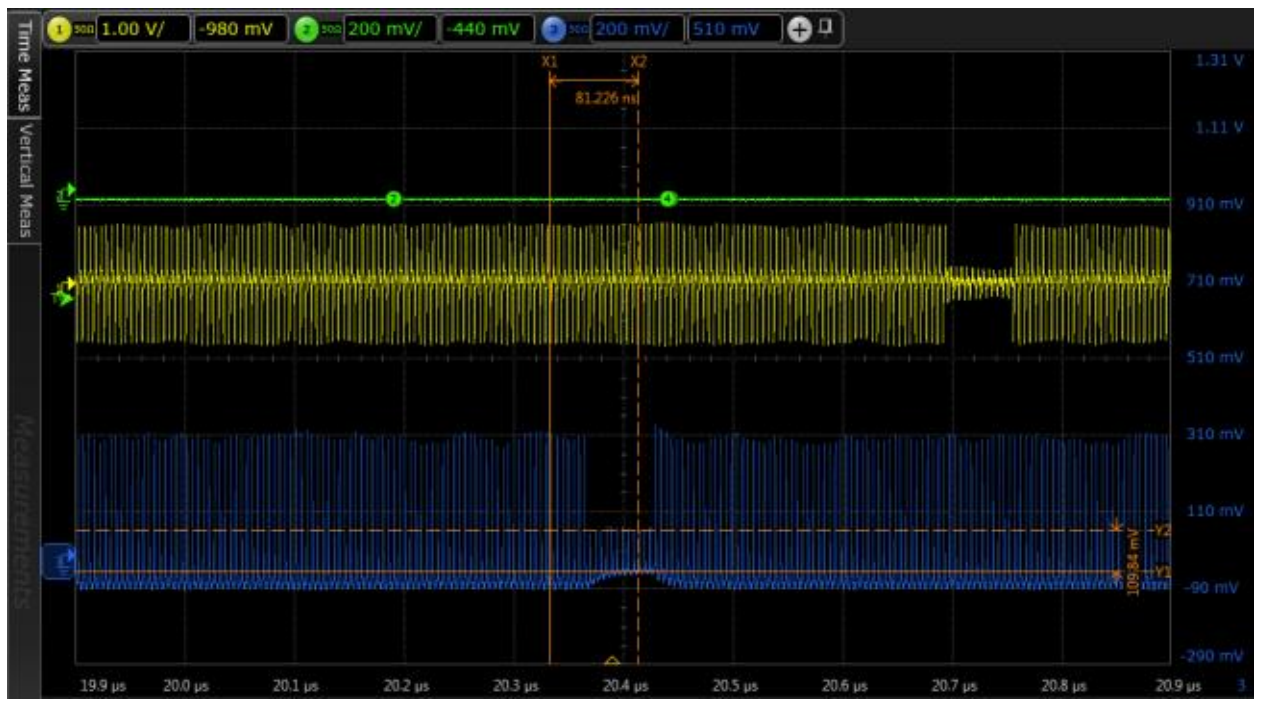

Figure 8: Expansion of the time base to show a single notch made up of 12 neutralized bunches on both detectors.

## *Comparison with estimation*

The laser total energy was varied from 47 mJ to 198 mJ for 13 notches each containing 12 pulses. The estimated 200 MHz pulse energy ranged from 0.12 to 0.5 mJ producing neutralizations from 14 to 73%. These are shown as data points in Figure 9. The measured laser parameters of pulse energy, transverse size, pulse length, and number of crossings were used with equations 1 & 2 to estimate the neutralization as a function of pulse energy. Figure 9 shows the estimate for both 21 (blue) and 23 (green) interactions.

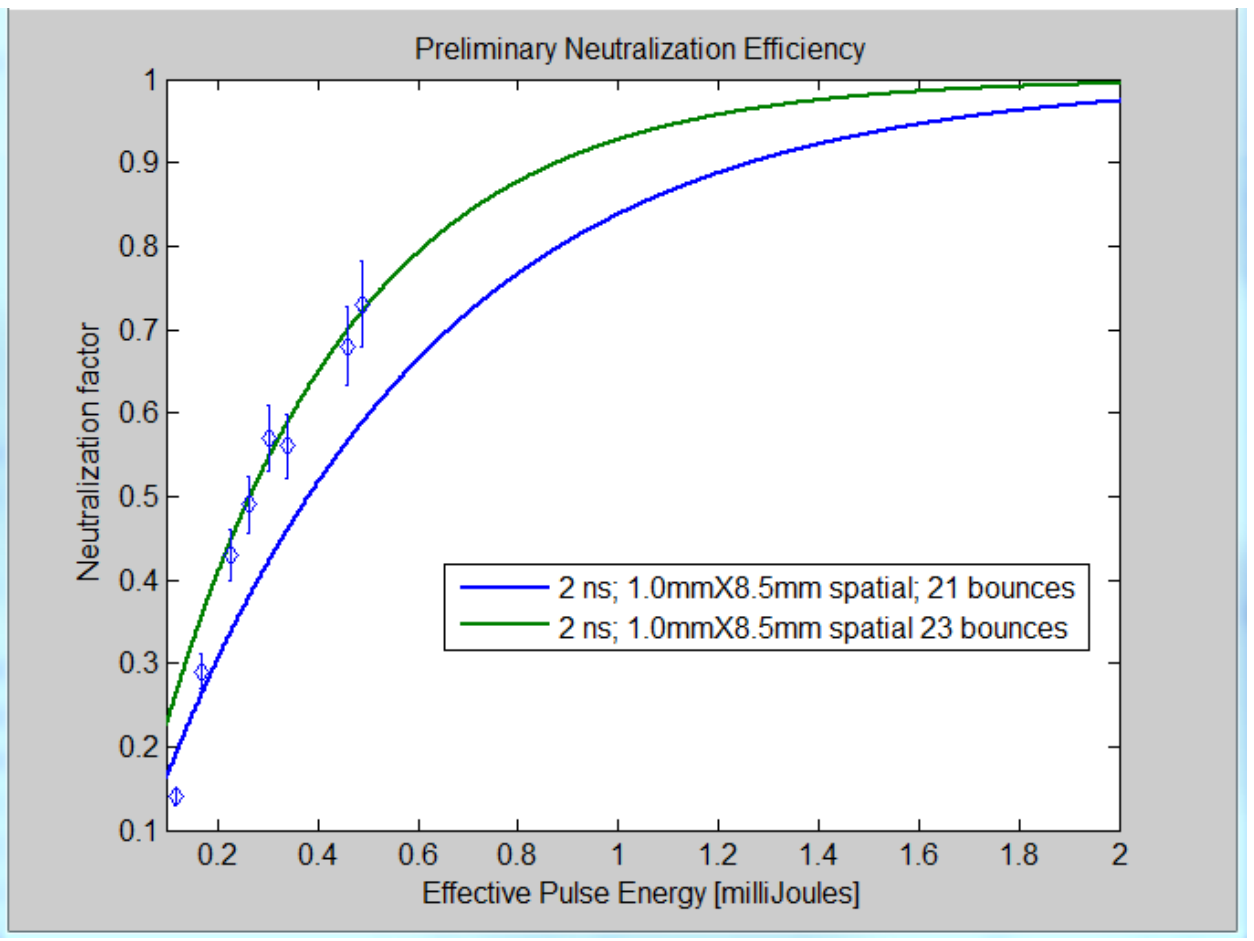

Figure 9: Comparison of measurements with estimates for 21 and 23 interactions.

## *Observation of notches in Booster*

Once notches were generated in the linac pulse, they were observed on an AC coupled phase monitor in the Booster ring. During the injection period, the RF cavities have about 35 kV of RF at the injection frequency of 38 MHz. Figure 10 shows the 38 MHz bunch structure during the injection period and the first turn after injection (top trace). The bottom trace shows the last injected turn and the first turn after injection. The depth of the notch is estimated by ratio of average offsets between the effected to non-effected bunch responses in the detector. Although this is a rough estimation, it is clear that the notches created in the linac pulse are reasonably preserved during the injection process.

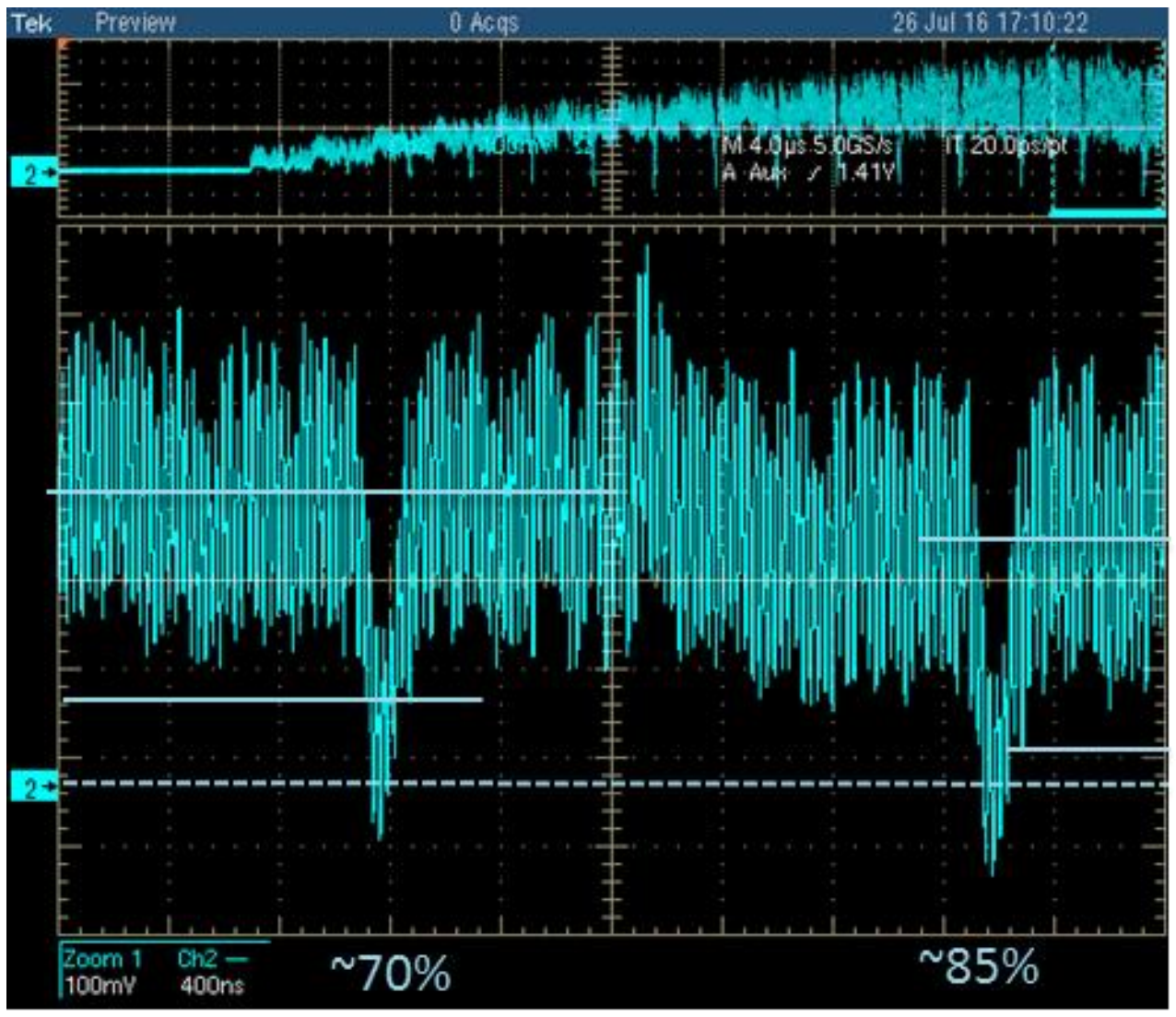

Figure 10: Booster phase monitor during injection.

Figure 11 shows the same signal several miliseconds after the completion of injection and RF adiabatic capture. There are two notches observed in the plot. The left notch was created in the ring while the notch on the right was created in the linac. During this study the creation of the notch in Booster had not been synchronized to that created in linac to investigate the survival of the linac notch. RF adiabatic capture currently takes place about 200 us after injection. During the period between last turn of injection and completion of capture, it is clear that beam has leaked into the notch created in linac. Simulations [7] show that the capture efficiency can be maintained and the notch remain clear at the level of 98% for an 80 ns notch if the start of capture is moved a few turns before the end of injection.

## NEAR TERM PLANS

There are several pieces of laser diagnostics and modifications for the fiber laser system that will be addressed to bring the pulse energy closer to the design value. Other items related to operational issues observed during the commissioning period will also be addressed. .In addition, the current notch generation in Booster will be synchronized to the linac notch and the details of adiabatic capture will be modified through the guidance of simulations.

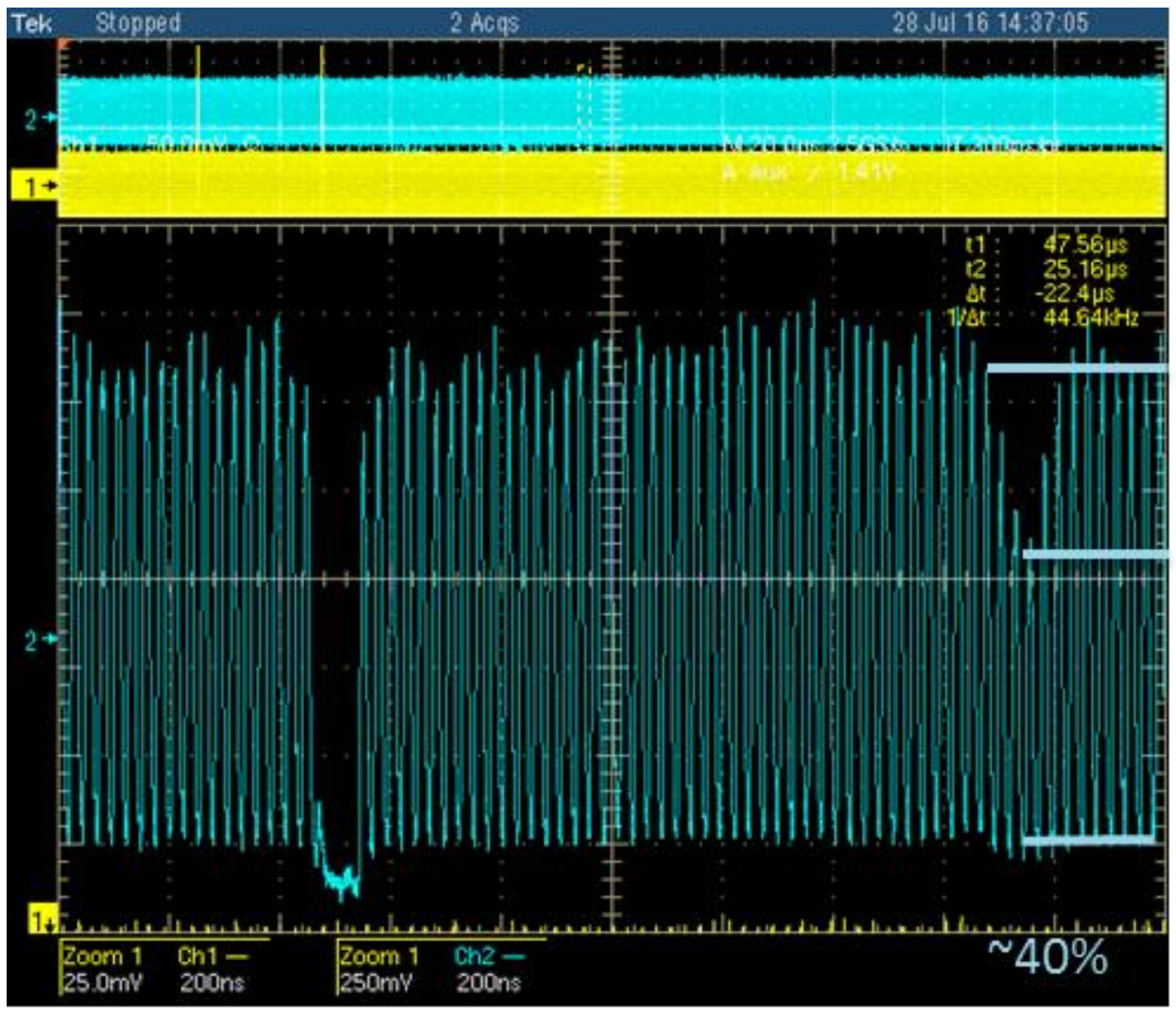

Figure 11: Booster phase monitor many milliseconds after injection.

## CONCLUSION

The Linac Laser Notcher was installed for the first time in the accelerator during a 2 week commissioning period at the end of July 2016. During this commissioning period we successfully demonstrated that the laser system was capable of creating multiple notches in the linac pulse that survived Booster injection.

## ACKNOWLEDGEMENT

We would like to acknowledge the many members of the Fermilab Accelerator Division their support and valuable help in this project: